\documentclass[pra,twocolumn,superscriptaddress,showpacs,preprintnumbers,amsmath,amssymb]{revtex4-1}

\usepackage{graphicx}
\usepackage{dcolumn}
\usepackage{bm}
\usepackage{ifpdf}

\begin{document}

\bibliographystyle{apsrev4-1} 


\title{Enhancement of spin coherence in a spin-1 Bose condensate by dynamical decoupling approaches}
\author{Bo-Yuan Ning}
\affiliation{Key Laboratory of Micro and Nano Photonic Structures (Ministry of Education), Department of Optical Science and Engineering, Fudan University, Shanghai 200433, China}%
\author{Jun Zhuang}
\affiliation{Key Laboratory of Micro and Nano Photonic Structures (Ministry of Education), Department of Optical Science and Engineering, Fudan University, Shanghai 200433, China}%
\author{J. Q. You}
\affiliation{Department of Physics, Fudan University, Shanghai 200433, China}%
\author{Wenxian Zhang}%
\email{wenxianzhang@fudan.edu.cn}
\affiliation{Key Laboratory of Micro and Nano Photonic Structures (Ministry of Education), Department of Optical Science and Engineering, Fudan University, Shanghai 200433, China}%

\date{\today}

\begin{abstract}
We study the enhancement of spin coherence with periodic, concatenated, or Uhrig dynamical decoupling $N$-pulse sequences in a spin-1 Bose condensate, where the intrinsic dynamical instability in such a ferromagnetically interacting condensate causes spin decoherence and eventually leads to a multiple spatial-domain structure or a spin texture. Our results show that all the three sequences successfully enhance the spin coherence by pushing the wave vector of the most unstable mode in the condensate to a larger value. Among the three sequences with the same number of pulses, the concatenated one shows the best performance in preserving the spin coherence. More interestingly, we find that all the three sequences exactly follow the same enhancement law, $k_- T^{1/2} = c$, with $k_-$ the wave vector of the most unstable mode, $T$ the sequence period, and $c$ a sequence-dependent constant. Such a law between $k_-$ and $T$ is also derived analytically for an attractive scalar Bose condensate subjecting to a periodic dynamical decoupling sequence.
\end{abstract}


\maketitle

\section{Introduction}

Spin coherence is of key importance to quantum devices utilizing spin degrees of freedom, e.g., a weak field magnetometer based on spin-1 Bose-Einstein condensates (BECs), which requires the spin coherence time to be as long as possible in order to improve the device sensitivity~\cite{Vengalattore07}. However, due to the ferromagnetic spin exchange interaction in the $^{87}$Rb Bose condensates, the intrinsic dynamical instability causes the unstable collective modes to grow exponentially into a spin-domain or spin-texture structure and eventually sabotages the spin coherence of the condensates~\cite{Chang05, Zhang2005PRL, Gu07, Vengalattore08, Kronjager10, Matuszewski10}. It is highly demanded to find a practical way to enhance the spin coherence in spinor Bose condensates in order to develop a higher precision magnetometer. In fact, the spin coherence is enhanced in a $^87$Rb spin-1 condensate by localizing the spin dynamics, which was proposed to unmask other interesting effects induced by the weak magnetic dipolar interactions~\cite{Zhang2010, Vengalattore08}.

Some experimental techniques may be employed to manipulate the atomic spin dynamics of a spinor condensate. One is the magnetic radio-frequency pulses, which can rotate the spin in an arbitrary angle along a certain direction~\cite{Bloch06,Vengalattore08}. Unfortunately, the spin exchange interaction in spin-1 condensates is rotationally invariant so that the magnetic pulses are not able to control the spin-exchange-interaction induced effects. The other is the optical Feshbach resonance technique, which can adjust both the sign and the magnitude of the spin exchange coefficient $c_2 = 4\pi\hbar^2 (a_2-a_0)/ 3m $ in a certain parameter range~\cite{Hamley09}, where $a_{0,2}$ is the $s$-wave scattering length and $m$ the mass of the atom. Theoretically, it has been shown that the dynamical instability of the $^{87}$Rb spin-1 condensate can be suppressed by periodically changing the spin exchange coefficient in a cosine form~\cite{Zhang2010}.

From the viewpoint of quantum control theory, other forms of modulating the spin exchange coefficient $c_2(t)$ are not only easy to realize in experiments but also powerful to suppress the spin decoherence. Recently, Uhrig proposed an optimal sequence with $N$ $\pi$-pulses to dynamically decouple a qubit from its boson environment in a spin-boson model~\cite{Uhrig07}. The Uhrig dynamical decoupling (UDD) sequence has also been applied in a system of a center spin coupled with many nuclear spins~\cite{Lee08, Yang08}. The results~\cite{Yang08} show that the UDD sequence can be used to remove the dephasing terms up to $\mathcal{O}(T^{N+1})$, where $T$ is the sequence period. This has aroused considerable experimental interests to demonstrate the advantages of UDD (see, e.g.,~\cite{Biercuk09, Du09, Lange10}). Moreover, the concatenated dynamical decoupling (CDD) sequence has also been demonstrated to be powerful in suppressing the decoherence in quantum systems~\cite{Khodjasteh05, Zhang07a, Yao07}. Despite of the success of these dynamical decoupling (DD) sequences, they have never been tested in a Bose-Einstein condensate, where the decoherence is dominated by a very different bath consisting of unstable collective modes.

In this paper, we theoretically investigate periodic dynamical decoupling (PDD), CDD, and UDD sequences on preserving the spin coherence of a homogenous $^{87}$Rb spin-1 Bose condensate. By comparing the performance of these sequences, we identify the optimal sequence in suppressing the dynamical instability of the spin-1 Bose condensate. Our results show that all sequences suppress the dynamic instability and the wave vector of the most unstable collective modes $k_-$ is pushed to a larger value. Although the CDD sequences behave better than the PDD or UDD sequences at the same number of pulses, the same law $k_{-}T^{1/2} = c$, with $c$ being a sequence-dependent constant, is followed for all the three sequences. Moreover, in a scalar Bose condensate, our analytical result for a PDD sequence further verifies this law between $k_{-}$ and $T$.

The paper is organized as follows. In Sec.~\ref{sec:dd}, we formulate the spin dynamics of a spin-1 condensate subjecting to a dynamical decoupling pulse sequence and describe our metric to measure the performance of the sequence. In Sec.~\ref{sec:num}, we present numerical results for a spin-1 condensate and compare the performances of PDD, UDD, and CDD sequences. To understand our numerical results, we analytically investigate, in Sec.~\ref{sec:scalar}, a simple case of a scalar condensate subjecting to a PDD sequence. The conclusion is given in Sec.~\ref{sec:con} and a brief derivation of our analytical results is provided in Appendix~\ref{sec:apd}.

\section{Spin dynamics of a spin-1 condensate subjecting to dynamical decoupling sequences}
\label{sec:dd}

In the mean-field theory, the evolution of a homogeneous three-component spin-1 Bose condensate is described by three coupled Gross-Pitaevski equations~\cite{Ho1998, Ohmi1998, Law1998}
\begin{eqnarray}
\label{Eq:1}
i\hbar\frac{\partial}{\partial t}\Phi_{\pm1} &=& [\mathcal{H}_0+c_{2}(n_{\pm1}+n_{0}-n_{\mp1})]\Phi_{\pm1}+c_{2}\Phi^{2}_{0}\Phi^{*}_{\mp1} \nonumber \\
i\hbar\frac{\partial}{\partial t}\Phi_{0} &=& [\mathcal{H}_0+c_{2}(n_{+}+n_{-})]\Phi_{0}+2c_{2}\Phi_{+1}\Phi^{*}_{0}\Phi_{-1},
\end{eqnarray}
where $\mathcal{H}_0=-(\hbar^{2}/2m)\nabla^{2}+V_{ext}+c_{0}n$ is the spin-independent Hamiltonian. In the above equations, $c_{0}=4\pi\hbar^{2}(a_{0}+2a_{2})/(3m)$ represents the spin-independent interaction and $n_i({\bf r})=|\Phi_i({\bf r})|^{2}$ is the total density of the condensate with $i=0,\pm 1$ and $\Phi_i$ the condensate wave function of the $i$th component.

The spin dynamics of the spin-1 condensate is determined by the rest spin exchange terms containing $c_2$. Depending on the sign of $c_2$, the spin exchange interaction can be either ferromagnetic ($c_2 < 0$, e.g., $^{87}$Rb condensates) or antiferromagnetic ($c_2 > 0$, e.g., $^{23}$Na condensates)~\cite{Ketterle1998,Stamper1998, Chapman2001, Sengstock04, Chang04, Kuwamoto04, Artur05,Lett07}. It was shown that a ferromagnetically interacting spin-1 Bose condensate is dynamically unstable while an antiferromagnetically interacting one is dynamically stable~\cite{Zhang2005PRL}. In other words, the spin coherence of a ferromagnetically interacting spin-1 condensate is lost, since some collective modes grow exponentially due to the intrinsic dynamical instability.

According to the Bogoliubov approximation, the time dependent equations of motion for the collective modes in a condensate can be cast in a matrix form as follows~\cite{Maldonado1993}:
\begin{equation}
\label{Eq:2}
i\hbar\frac{\partial{{\bf x}}}{\partial{t}}=M\cdot{\bf x},
\end{equation}
where ${\bf x} = (\delta\Psi_{+1},\delta\Psi_{0}, \delta\Psi_{-1}, \delta\Psi^{*}_{+1}, \delta\Psi^{*}_{0},\delta\Psi^{*}_{-1})^T$ denotes the amplitude of the collective modes and $M$ is an effective Hamiltonian. In general, $M$ has a matrix form as
\begin{equation}
\label{Eq:3}
M=\left(\begin{array}{cc}A & B \\-B^{*} & -A^{*}\end{array}\right),
\end{equation}
where for a spin-1 condensate
\begin{eqnarray}
A &=& \varepsilon_{k}I+c_{0}A_{0}+c_{2}A_{2}, \nonumber \\
B &=& c_{0}B_{0}+c_{2}B_{2}. \nonumber
\end{eqnarray}
Here, $\varepsilon_{k} = \hbar^2 k^2 / (2m)$ is the kinetic energy of the collective mode in a plane wave form with a wave vector $k$, $I$ is a $3\times3$ identity matrix, and $A_{0,2}, B_{0,2}$ are also $3\times3$ matrices given by
\begin{eqnarray}
 A_0 &=& \left(
               \begin{array}{ccc} n_{+1} & \Phi_{0}^{*}\Phi_{+1} & \Phi_{-1}^{*}\Phi_{+1} \\
                                  \Phi_{0}\Phi_{+1}^{*} & n_{0} & \Phi_{0}\Phi_{-1}^{*}\\
                                  \Phi_{-1}\Phi_{+1}^{*} & \Phi_{0}^{*}\Phi_{-1} & n_{-1}\\
               \end{array}\right), \nonumber \\
 A_2 &=& \left(
               \begin{array}{ccc} n_{+1}+n_{0} & \Phi_{0}\Phi_{-1}^{*} & -\Phi_{-1}^{*}\Phi_{+1} \\
                                  \Phi_{0}^{*}\Phi_{-1} & n_{+1}+n_{-1} & \Phi_{0}^{*}\Phi_{+1}\\
                                  -\Phi_{-1}\Phi_{+1}^{*} & \Phi_{0}\Phi_{+1}^{*} & n_{-1}+n_{0}\\
               \end{array}\right), \nonumber \\
 B_0 &=& \left(
               \begin{array}{ccc} n_{+1} & \Phi_{0}\Phi_{+1} & \Phi_{-1}\Phi_{+1} \\
                                  \Phi_{0}\Phi_{+1} & n_{0} & \Phi_{0}\Phi_{-1}\\
                                  \Phi_{-1}\Phi_{+1} & \Phi_{0}\Phi_{-1} & n_{-1}\\
               \end{array}\right), \nonumber \\
 B_2 &=& \left(
               \begin{array}{ccc} n_{+1} & \Phi_{0}\Phi_{+1} & n_{0}-\Phi_{-1}\Phi_{+1} \\
                                  \Phi_{0}\Phi_{+1} & 2\Phi_{-}\Phi_{+} & \Phi_{0}\Phi_{-1}\\
                                  n_{0}-\Phi_{-1}\Phi_{+1} & \Phi_{0}\Phi_{-1} & n_{-1}\\
               \end{array}\right). \nonumber
\end{eqnarray}
Obviously, matrix $M$ is neither Hermitian nor symmetric although $A_{0,2}$ are Hermitian and $B_{0,2}$ are symmetric. However, the matrix $M$ has paired eigenvalues with the same amplitude but different sign, due to its particular form~\cite{Maldonado1993}. The dynamical instability of a condensate is manifested as the existence of at least one pair of complex eigenvalues~\cite{Zhang2005PRL}.

By employing the optical Feshbach resonance technique, the spin exchange coefficient $c_{2}$ can be tuned not only in amplitude but also in sign~\cite{Hamley09}. It occurs naturally to utilize such a technique to suppress the dynamical instability in a ferromagnetically interacting spin-1 Bose condensate. Previous theoretical investigation has indeed demonstrated the suppression effect by modulating $c_2(t)$ in a simple cosine form~\cite{Zhang2010}.

From the viewpoint of the control theory, there are many more complicated DD pulse sequences which may outperform the simple cosine modulation of $c_2(t)$ in suppressing the dynamical instability and enhancing the spin coherence. In addition, these pulse sequences are easier to implement in experiments. Hereafter, applying a dynamical decoupling pulse means alternating the sign of $c_2$ but keeping its amplitude unchanged.

\begin{figure}
\includegraphics[width=3.25in]{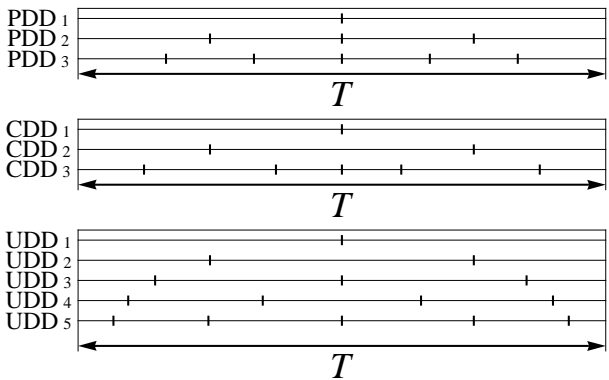}
\caption{Various PDD (top), CDD (middle), and UDD (bottom) sequences. The vertical short lines mark the time of each applied pulse.}
\label{Fig:1}
\end{figure}

A PDD sequence denotes flipping the sign of $c_2$ back and forth periodically [see top panel in Fig.~\ref{Fig:1}]. By concatenating PDD sequences at different levels, we obtain the CDD sequences [see middle panel in Fig.~\ref{Fig:1}]. A UDD sequence [see bottom panel in Fig.~\ref{Fig:1}] containing $N$ pulses is defined as~\cite{Uhrig07, Lee08, Yang08}
\begin{equation}
\label{Eq:4}
\tau_{j}=\frac{T}{2} \left[\cos\left(\frac{\pi(j-1)}{N+1}\right) - \cos\left(\frac{\pi{j}}{N+1}\right)\right],
\end{equation}
where $\tau_{j}$ is the delay between the $(j-1)$th and $j$th pulses and $T=\sum_{j=1}^{N+1}\tau_{j}$ is the period of the total sequence. We notice that the CDD$_1$ (CDD at one-level concatenation), UDD$_1$ (one-pulse UDD), and PDD sequence are the same, and CDD$_2$ and UDD$_2$ are also the same.

There are various choices of metric to measure the pulse sequence effect. We choose the wave vector of the most unstable mode, $k_-$, since this mode dominates in the dynamics of the spin-1 condensate~\cite{Zhang2005PRL}. The larger the value of $k_-$ is, the better the performance of the pulse sequence will be. For a spin-1 condensate under many DD pulses, we adopt a numerical method (se below) to obtain the $k_-$.

By applying a sequence of $N$ DD pulses, the time evolution operator $U_{T}$ of the collective modes of the spin-1 Bose condensate is
\begin{eqnarray}
U_{T} &\equiv & {\mathcal{T}}\exp\left[-\frac{i}{\hbar}\int^{T}_{0} dt' M(t')\right] \nonumber \\
 & = & \exp[-iM_{\pm}\tau_{N+1}] \ldots\exp[-iM_{\mp}\tau_{2}] \exp[-iM_{\pm}\tau_{1}], \nonumber \\
\label{Eq:5}
\end{eqnarray}
where ${\mathcal{T}}$ denotes the time-ordering operator, and $M_{\pm}$ denotes the $M$ matrix with $c_{2}$ being either positive or negative. By diagonalizing $U_T$ and reexpressing it as $U_{T}=V^{\dagger}\exp(-iTF)V$ with $F$ a $6\times 6$ diagonal matrix, we identify the wave vector of the most unstable mode $k_-$ by searching the largest imaginary part of $F$ for different wave vector $k$~\cite{Zhang2010}.

\section{Numerical results for a spin-1 Bose condensate}
\label{sec:num}

\begin{figure}
\includegraphics[width=3.4in]{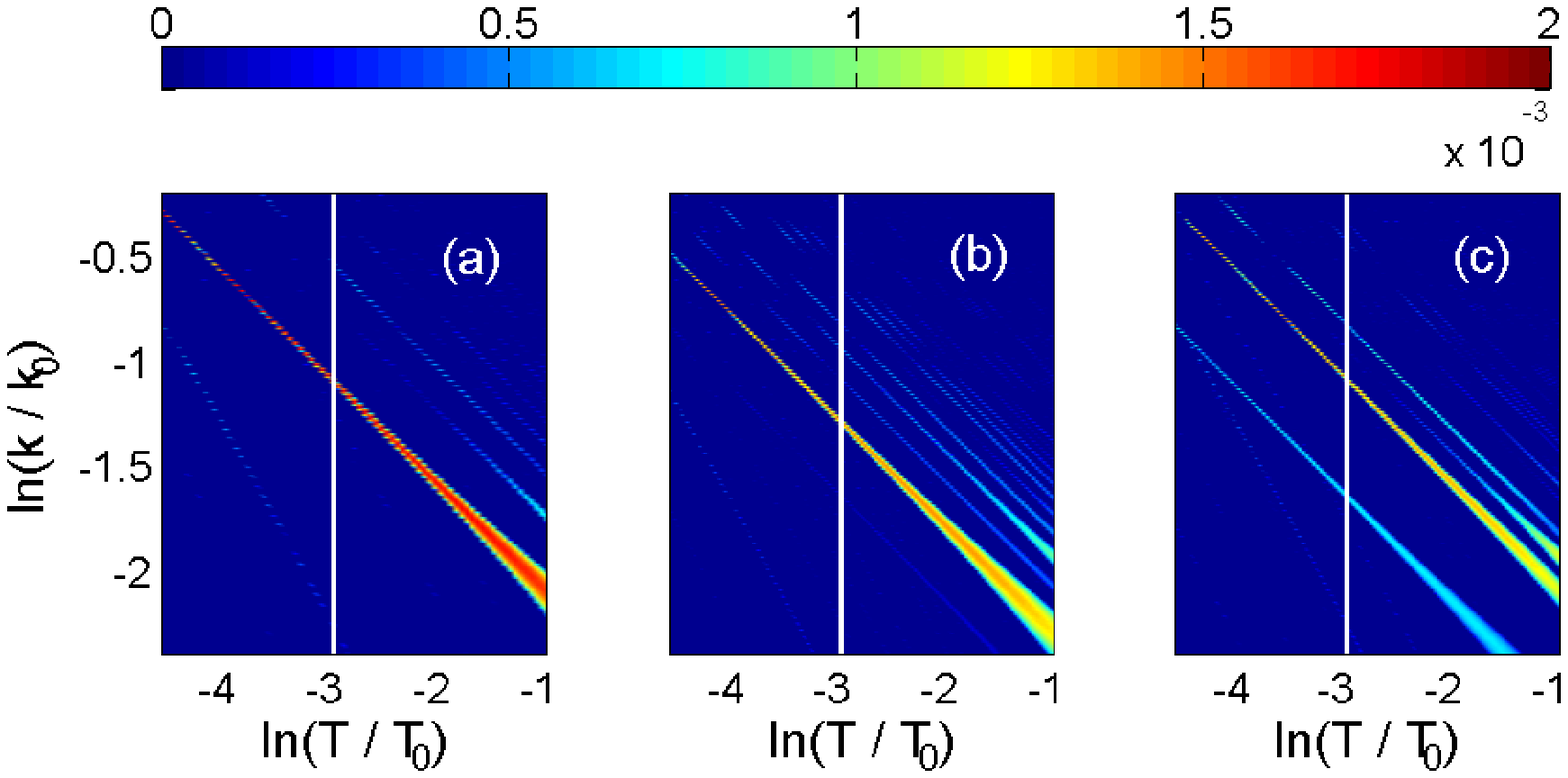}
\includegraphics[width=3.3in]{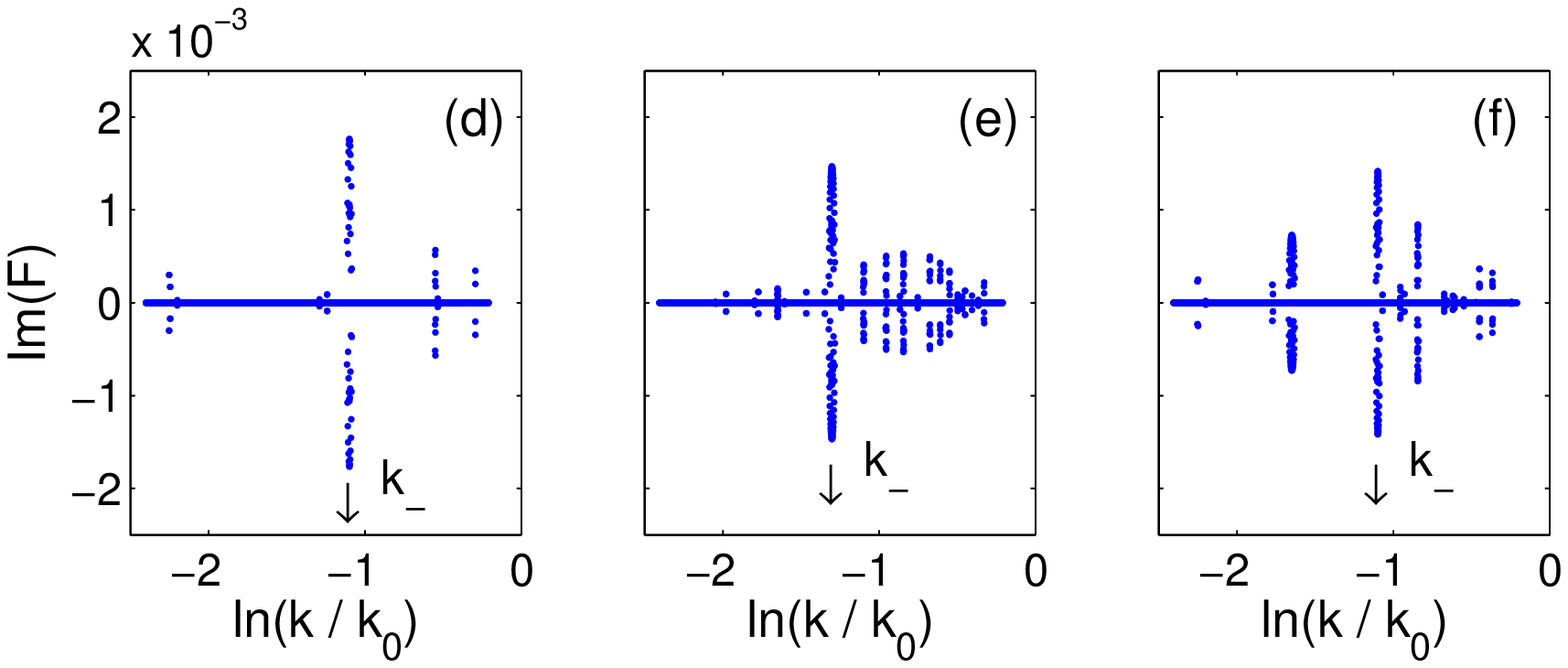}
\caption{(Color online) Dependence of the imaginary part of the Floquet operator $F$ on the wave vector $k$ and the period $T$ for sequences (a) PDD$_3$, (b) UDD$_5$, and (c) CDD$_3$. Lower panels show the $k$ dependence for (d) PDD$_3$, (e) UDD$_5$, and (f) CDD$_3$ at $T=T_{0}/20$ which is marked by the vertical white lines in panel (a), (b), and (c), respectively. Here, we use parameters~\cite{Pu1999,Zhang2010} $k_0=1$ and $T_0=\pi/\left|c_{2}n\sqrt{n_{0}(1-n_{0})}\cos(\theta_{0}/2)\right|=\pi/(0.4|c_2|n)$, where the initial condition is $\theta_{0}=0$ and $n_0= 0.8$. In calculations we have set $c_0=1$, $n=1$, and $c_2=-0.0046$ which are the parameters of $^{87}$Rb spin-1 condensates.}
\label{Fig:2}
\end{figure}

Following the procedure described at the end of Sec.~\ref{sec:dd}, we numerically obtain the dependence of the unstable modes on the DD sequence periods for a spin-1 Bose condensate. For the numerical simplicity, we set $m=1$ and $\hbar=1$. The results are presented in Fig.~\ref{Fig:2} and Fig.~\ref{Fig:3}(a). Several interesting observations are listed as follows:

\newcounter{Lcount}
\begin{list}{\roman{Lcount})}
{\usecounter{Lcount}}
\item Compared to the free evolution ($T=T_0$), all the DD sequences suppress the dynamical instability of the spin-1 condensate in three ways, narrowing the area of the unstable region, shifting $k_-$ to a larger value, and decreasing the value of $F_-=Im(F)|_{k_-}$ (the exponential growth rate of the most unstable mode).
\item Comparing the performances of the same DD sequence at different period $T$, $F_-$ is almost the same but $k_-$ is shifted to a larger value when decreasing $T$.
\item At the same number of pulses ($N=5$) and the same sequence period, the UDD sequence shifts $k_-$ the least, while the CDD and PDD sequences behave almost the same (but both better than the UDD) [see Fig.~\ref{Fig:3}(a)]. Moreover the CDD sequence suppresses the growth rate $F_-$ more than the PDD [see Figs.~\ref{Fig:2}(d) and ~\ref{Fig:2}(f)].
\item The relationship between $k_-$ and $T$ for all PDD, UDD, and CDD sequences follows exactly the same power law: $k_- T^{1/2} = c$, where $c$ is a sequence-dependent constant and can be obtained by numerically fitting the slope of the lines in Fig.~\ref{Fig:3}(a)\label{lcout4}.
\end{list}

The above results clearly indicate that the dynamical instability in a ferromagnetically interacting spin-1 Bose condensate is essentially inhibited once the wavelength of the most unstable mode $2\pi/k_-$ is less than the spin healing length $\xi=h/\sqrt{2m|c_2|n}$. Our results shows that CDD sequences most effectively suppress the dynamical instability  and they are the best choice to enhance the spin coherence in a spin-1 Bose condensate.

It is not easy to fully understand all the above numerical results for a spin-1 Bose condensate in a simple picture, especially the result in (\romannumeral4), because it is difficult to obtain analytical results with either the effective Hamiltonian $M$ matrix [Eq.~(\ref{Eq:3})] or the $F$ matrix. To circumvent these difficulties, we investigate the effect of the same DD sequences but in a scalar Bose condensate, where it becomes possible for us to obtain the same behaviors but in an analytical way.

\section{results for scalar Bose condensate}
\label{sec:scalar}

Similar to the case of a spin-1 Bose condensate, we also focus on the change of $k_{-}$ for the most unstable mode of a scalar condensate subjecting to DD pulses, which alternates the sign of the interaction coefficient $g=4\pi\hbar^2 a_s/m$, with $a_s$ the $s$-wave scattering length. The interaction between atoms is repulsive if $g>0$ and attractive if $g<0$~\cite{Dalfovo}. For a homogeneous scalar condensate with attractive interaction, the system is intrinsically unstable~\cite{Kagan}. By  periodically flipping the sign of $g$, it is possible to suppress the instability of an attractively interacting Bose condensate~\cite{Zhang2010}. Fortunately, both the sign and the amplitude of $g$ can be precisely tuned in experiments via the Feshbach resonance technique~\cite{Roberts1998,Roberts2001}.

The time-dependent effective Hamiltonian $M(t)$ matrix of the collective modes in the scalar condensate modulated by a PDD$_1$ sequence is given by
\begin{equation}
M(t) =\left\{\begin{array}{cc}
           M_+, & t \in \left[0, \frac T 2\right), \\
           M_-, & t \in \left[\frac T 2, T\right),
         \end{array} \right. \nonumber
\end{equation}
where
\begin{equation}
M_{\pm }=\left(\begin{array}{cc}
           \varepsilon_{k}\pm gn & \pm gn \\
           \mp gn & -(\varepsilon_{k}\pm gn)
         \end{array}\right).
         \label{Eq:6}
\end{equation}
Following exactly the same procedures in Eq.~(\ref{Eq:5}) for the above $M$ matrix, we first diagonalize the corresponding time evolution operator $U_T$, then find out the eigenvalues of the Flouqet operator $F$, and finally extract the mode containing largest imaginary part $F_-$ and the corresponding wave vector $k_-$. For a scalar Bose condensate, the matrix $M(t)$ is much simpler than the one in a spin-1 condensate, so it is possible for us to obtain analytical results for a PDD$_1$ sequence. For convenience, we reform the matrix $M_\pm$ in Pauli matrices as
\begin{equation}
M_{\pm}=(\varepsilon_{k}\pm gn)\cdot\sigma_{z}\pm ign\cdot\sigma_{y}.
\label{Eq:7}
\end{equation}
By utilizing the properties of Pauli matrices, it is straightforward to obtain the time evolution operator
\begin{eqnarray}
U_{\pm} &\equiv & \exp(-i\tau M_{\pm}) \nonumber \\
    &=& \cos(\tau{x_{\pm}}) - \frac{i}{x_{\pm}} [(\varepsilon_{k}\pm{gn})\cdot\sigma_{z}\pm ign\cdot\sigma_{y}] \sin(\tau x_{\pm}), \nonumber
\end{eqnarray}
where $x_{\pm}=\sqrt{\varepsilon_{k}(\varepsilon_{k}\pm 2gn)}\;$ and $\tau=T/2$. The propagator in a complete period is
\begin{equation}
U_{T}=U_{-}U_{+}=\left(\begin{array}{cc}P-iQ_{z} & -Q_{x}+iQ_{y}\\
-Q_{x}+iQ_{y} & P+iQ_{z} \end{array}\right),
\label{Eq:8}
\end{equation}
and the corresponding eigenvalues of $U_{T}$ is
\begin{equation}
\lambda=P\pm\sqrt{Q_{x}^{2}+Q_{y}^{2}-Q_{z}^{2}},
\label{Eq:9}
\end{equation}
where
\begin{eqnarray*}
P&=&\cos(\tau x_{-})\cos(\tau x_{+}) - \frac{\varepsilon_{k}^{2}}{x_{-}x_{+}} \sin(\tau x_{-}) \sin(\tau x_{+}), \\
Q_{x}&=&\frac{2\varepsilon_{k}gn}{x_{-}x_{+}} \sin(\tau x_{-})\sin(\tau x_{+}),\\
Q_{y}&=&\frac{gn}{x{-}}\sin(\tau x_{-})\cos(\tau x_{+}) - \frac{gn}{x_{+}} \cos(\tau x_{-})\sin(\tau x_{+}),\\
Q_{z}&=&\frac{\varepsilon_{k}-gn}{x_{-}} \sin(\tau x_{-})\cos(\tau x_{+}) \\
    && + \frac{\varepsilon_{k}+gn}{x_{+}} \cos(\tau x_{-})\sin(\tau x_{+}).
\end{eqnarray*}
Since $\lambda=\exp(-iTF)$, the imaginary part of the Flouquet operator follows as
\begin{equation}
Im(F)=\frac{1}{T}\ln{|\lambda|}.\label{Eq:10}
\end{equation}

\begin{figure}
\includegraphics[width=3.25in]{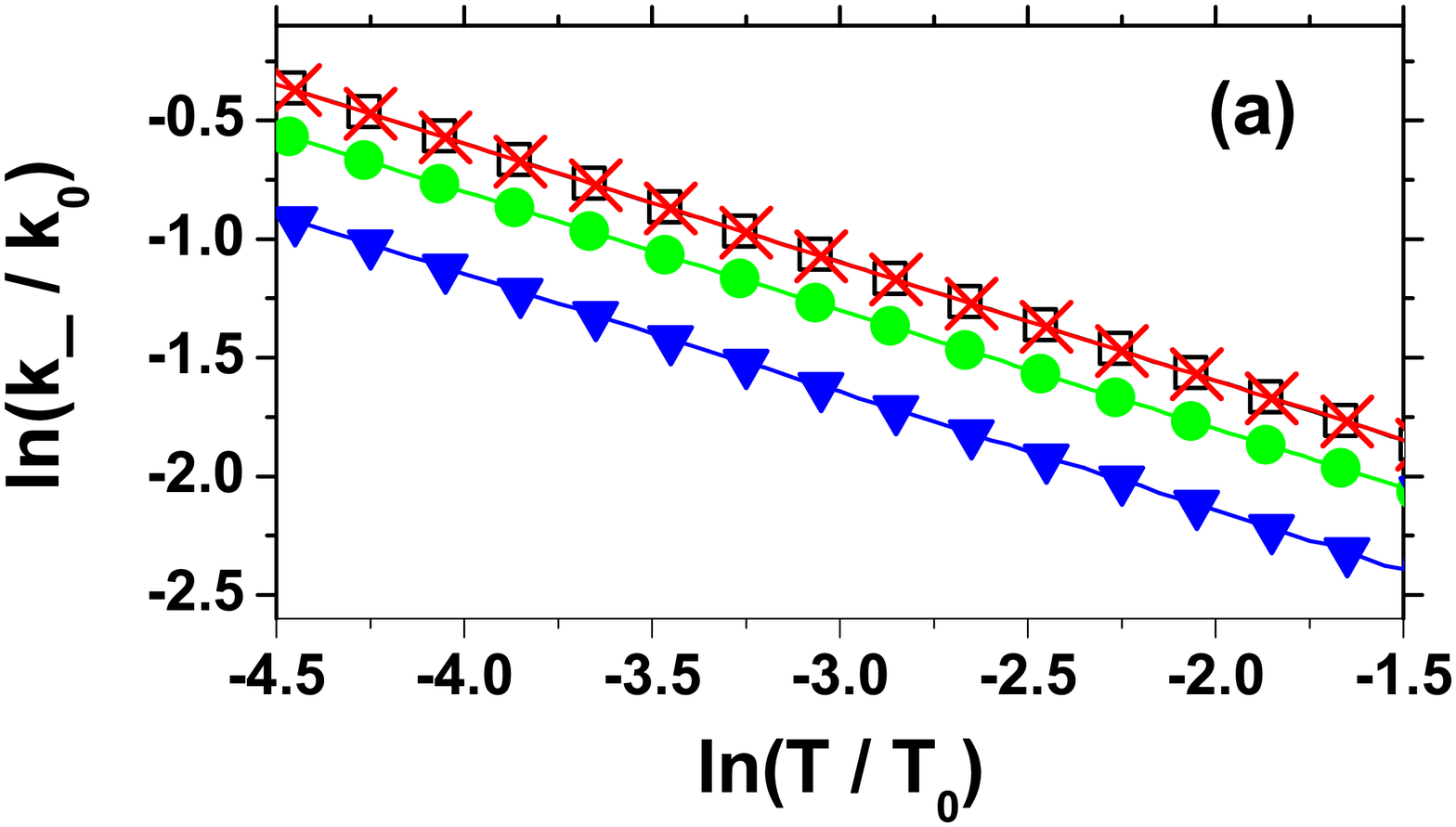}
\includegraphics[width=3.25in]{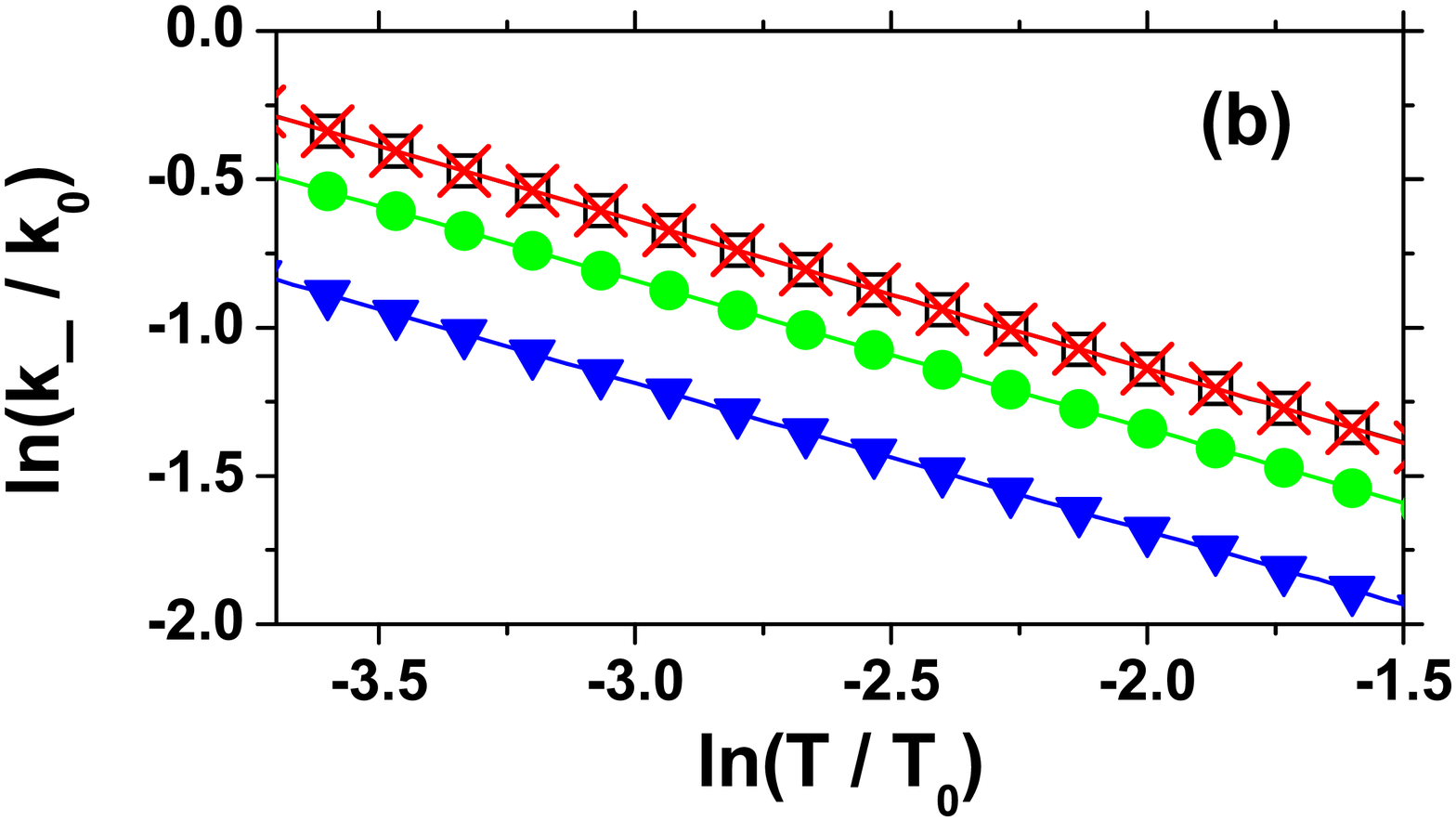}
\caption[]{\label{Fig:3}(Color online) Numerical results of power law relationship between the wave vector $k_-$ of the most unstable mode and the sequence period $T$ for (a) spin-1 condensates and (b) scalar condensates: open squares --- CDD$_3$, crosses --- PDD$_3$, filled circles --- UDD$_5$, filled triangles --- PDD$_1$. The specific fitting results are shown in Table~\ref{tab1}. The parameters for the scalar condensates are $g=-0.0046$ and $n=1$.}
\end{figure}

We notice that the kinetic energy $\varepsilon_k$ of the most unstable mode is much larger than the interaction energy $gn$ if the PDD$_1$ sequence period $T$ is small. By assuming $gn \ll \varepsilon_{k}$ and keeping to the second order of $gn / \varepsilon_{k}$, we analytically obtain the relationship between the wave vector $k_{-}$ of the most unstable mode and the sequence period $T$ from Eqs.~(\ref{Eq:8})-~(\ref{Eq:10}) as
\begin{equation}
k_{-} T^{1/2} = c,
\label{Eq:11}
\end{equation}
with $c=\sqrt{2\pi}$. For a detailed derivation of Eq.~(\ref{Eq:11}), see Appendix~\ref{sec:apd}.

\begin{table}
\caption{\label{tab1}Numerical fitting values of the slope, intercept and calculated $c^2/2\pi$ for lines in Fig.~\ref{Fig:3} for spin-1 condensates [denoted below as (a)] and scalar condensates [denoted as (b)]. In calculations, $c^2 = \pi \exp(2y_0) /\left[|c_2|n\sqrt{n_0(1-n_0)}\right]$ for spin-1 condensates and $c^2 = 2\pi \exp(2y_0)/(|g|n)$ for scalar condensates.}
\begin{ruledtabular}
\begin{tabular}{cccc}
 & slope & intercept ($y_0$) & $c^2/2\pi$ \\
 \colrule
 (a) PDD$_1$& -0.473 & -3.051 & 0.608\\
 (a) UDD$_5$& -0.499 & -2.799 & 1.006\\
 (a) PDD$_3$& -0.495 & -2.583 & 1.551\\
 (a) CDD$_3$& -0.495 & -2.583 & 1.551\\
 \toprule
 (b) PDD$_1$& -0.498 & -2.681 & 1.020\\
 (b) UDD$_5$& -0.499 & -2.341 & 2.013\\
 (b) PDD$_3$& -0.499 & -2.137 & 3.027\\
 (b) CDD$_3$& -0.499 & -2.137 & 3.027\\

\end{tabular}
\end{ruledtabular}
\end{table}

It is difficult to obtain the analytical results for other more complex sequences such as UDD$_5$, PDD$_3$ and CDD$_3$, so we carry out numerical calculations for the scalar condensates. The results are shown in Fig.~\ref{Fig:3}(b). It is clearly shown in Fig.~\ref{Fig:3} that no matter in a spin-1 condensate or in a scalar condensate, all sequences exhibit the same relationship between $k_-$ and $T$: $k_{-} T^{1/2} = c$, with $c$ a sequence-dependent constant (see Table~\ref{tab1}).

\section{Conclusion}
\label{sec:con}

In conclusion, we have theoretically investigated the PDD, CDD and UDD pulse sequences on suppressing the unstable collective modes in a spin-1 Bose Einstein condensate. Our numerical results show that the three DD sequences can be used to effectively protect the coherence of the spin-1 condensate. The wave vector $k_-$ of the most unstable mode is shifted to a much larger value, which in fact enhances the spin coherence of the condensate. Furthermore, using the analytical results in a scalar condensate, we reveal an interesting relation between $k_{-}$ and sequence period $T$: $k_- T^{1/2} = c$, where $c$ is a sequence-dependent constant.

\acknowledgments

BYN and WZ acknowledge support by the National Natural Science Foundation of China Grant No. 10904017, NCET, Specialized Research Fund for the Doctoral Program of Higher Education of China under Grant No. 20090071120013, and Shanghai Pujiang Program under Grant No. 10PJ1401300. JQY and WZ acknowledge support by National Basic Research Program of China Grant No. 2009CB929300.

\appendix
\section{Derivation of Eq.~(\ref{Eq:11})}
\label{sec:apd}

According to the discussion in Sec.~\ref{sec:scalar}, we extract the most unstable mode which has the largest growth rates $F_-$, satisfying $d[Im(F)]/dk|_{k_-}=0$. Using Eq.~(\ref{Eq:10}), we find that the derivative of $F$ is
\begin{equation}
\frac{dIm(F)}{dk}=\frac{1}{T}\frac{d(\ln{|\lambda|})}{dk}, \label{Eq:A1}
\end{equation}
where
\begin{equation}
|\lambda|=\left \{ \begin{array}{cl}
                                      \left|P+\sqrt{Q_{x}^{2}+Q_{y}^{2}-Q_{z}^{2}}\right|, & \lambda \; {\rm is \; real}, \\
                                      \sqrt{P^{2}+Q_{z}^2-Q_{x}^2-Q_{y}^2}, & \lambda \; {\rm is \; complex},
                              \end{array} \right. \label{Eq:A2}
\end{equation}
with $P$ and $Q_{x,y,z}$ given in Eq.~(\ref{Eq:9}). In the complex $\lambda$ case, $|\lambda|= 1$ and $Im(F)=0$, which means a stable collective mode. While in the real $\lambda$ case, by defining $\epsilon=gn/\varepsilon_{k} \ll 1$, $\lambda$ is reformed as
\begin{widetext}
\begin{equation}
\label{Eq:A4}
\begin{split}
\lambda=&\cos{y_+}\cos{y_-}-\frac{1}{\sqrt{1-4\epsilon^{2}}}\sin{y_+}\sin{y_-}+ \\
        &\sqrt{\frac{1}{8\epsilon^2-2}\left[1 - \cos{2y_+}\cos{2y_-} + 2\epsilon^2\left(\cos{2y_{+}}+\cos{2y_{-}} + \cos{2y_+}\cos{2y_-} - 3\right) + \sqrt{1-4\epsilon^{2}}\sin{2y_+}\sin{2y_-}\right]}\;\;,
\end{split}
\end{equation}
where $y_{\pm}=\varepsilon_{k}\tau\sqrt{1\pm2\epsilon}$. By keeping terms up to $\epsilon^2$, we obtain
\begin{eqnarray}
\label{Eq:A5}
\lambda & \approx & \cos(y_{+}+y_{-})-2\epsilon^2\sin{y_{+}}\sin{y_{-}}
     + \sqrt{-\sin^2(y_{+}+y_{-})-4\epsilon^2\cos (y_{+}+y_-)\sin{y_+}\sin{y_-}}\;\;.
\end{eqnarray}
\end{widetext}
Considering the fact that $\epsilon \ll 1$ and $\lambda$ is real (which requires the terms under the square root to be of the order of $\epsilon^2$), we have $y_{+}\approx y_{-}\approx k^2\tau / 2$, and
\begin{equation}
\label{Eq:A6}
\lambda \approx \cos(y_{+}+y_{-}) + O(\epsilon).
\end{equation}
Thus, as long as $y_{+}+y_{-}=\pi$,
$$\frac{dF_-}{dk}=0,$$
which gives rise to the relationship in Eq.~(\ref{Eq:11}), i.e., $k_{-}T^{1/2} = \sqrt {2\pi}$.

Another concern is the suppression of $F_-$. According to Eq.~(\ref{Eq:A5}), when $k_- T^{1/2} = \sqrt {2\pi}$,
\begin{equation}
\label{Eq:A7}
\lambda=-1+4\frac{|g|n}{k_-^2}
\end{equation}
if we keep terms up to the order of $\epsilon$. Finally, we obtain that
\begin{equation}
\label{Eq:A8}
F_- = \frac{1}{T}\ln\left(1-4\frac{|g|n}{k_-^2}\right),
\end{equation}
which shows that $F_-$ depends very weakly on $k_-$.

\end{document}